# Time to Quantify Falsifiability

Ilya Nemenman[1]

Departments of Physics and Biology, Emory University,
Atlanta, GA 30322, USA

May 29, 2015

Abstract: Here we argue that the notion of falsifiability, a key concept in defining a valid scientific theory, can be quantified using Bayesian Model Selection, which is a standard tool in modern statistics. This relates falsifiability to the quantitative version of the Occam's razor, and allows transforming some long-running arguments about validity of certain scientific theories from philosophical discussions to mathematical calculations.

Reinvigorated by Steinhardt's criticism of unfalsifiability of multiverse inflationary cosmology,[1] which was flexible enough to explain both negative *and* positive results of the BICEP2 experiment,[2] the past year has resulted in a renewed interest in the old debate: What defines the scientific method?[3] What makes a good physical theory? While the underlying inflationary theory is mathematically sophisticated and modern, the debate itself has been surprisingly qualitative[3], similar to what it could have been long ago, when Popper brought falsifiability in the spotlight.[4] Such data-less and often extremistically binary arguments (e.g., should falsifiability be "retired" altogether?[5]) seem out of place in data-driven, real-valued scientific word.

In fact, we already have a mathematical framework to do much better! It is based on statistical principles that have long been a part of science. In particular, falsifiability is not an independent concept: its graded, real-valued generalization emerges automatically from the empirical nature of science, similarly to how the Occam's razor transformed itself from a qualitative philosophical principle into a statistical result.[6,7]

This is easy to see in the language of Bayesian statistics. Suppose we want to decide which of two theories, $T_1$ and $T_2$, explains the world better. Our *a priori* knowledge of this is summarized in Bayesian priors, $P_{1,2}$. After experimental data **x** are collected, the ratio of posterior probabilities of the theories is given by the Bayes theorem, $\frac{P(T_1|x)}{P(T_2|x)} = \frac{P(x|T_1)P_1}{P(x|T_2)P_2}$, where $P(x|T_{1,2})$ are the likelihood terms, that is, the probabilities to get the observed data within the theory. The likelihood increases when the theory "fits" the data. However, because probabilities must be normalized, the likelihood decreases as the inverse of the total number of all typical data sets that could have been generated within the theory. This tradeoff between the *quality of fit* and the *statistical complexity* is known as Bayesian model selection, and it is used routinely in modern statistics. It provides an *automatic* Occam razor against statistically complex theories, which depends only weakly on specifics of the priors.

At an extreme, *any* data set is equally compatible with an unfalsifiable theory, and hence can come from it with the same probability. Thus the likelihood is equal to the inverse of the total possible number of experimentally distinct data sets. In contrast, a falsifiable theory would be incompatible with some data, and hence would have a higher probability of generating the other data. The difference between the theories would grow with the number of conducted experiments. Thus within Bayesian model selection, any falsifiable theory that fits data well would win eventually, unless the unfalsifiable theory had astronomically higher *a priori* odds. For example, evolution cannot generate "fossil rabbits in the Precambrian" (J.B.S. Haldane). This leads to an immediate *empirical*, *quantitative* choice of evolutionary theory over creationism as the best explanation of the fossil record, without the need to reject creationism *a priori* as unscientific.

---

[1] Email: ilya.nemenman@emory.edu

In other words, there is no need to require falsifiability of physical theories: it follows directly from statistical principles, on which empirical science is built. Its statistical version is more nuanced, as has been recognized by philosophers.[7] The practical applications are hard, requiring computing probabilities of arbitrary experimental outcomes (in fact, it was an error in such a computation that started the current controversy). In addition, there is an uncomfortable possibility that statistics can reject a *true* theory, which just happens to be unfalsifiable. And yet, crucially, statistical model selection is *quantitative* and *evidence-driven*, potentially moving the inflationary multiverse debate and similar discussions from the realm of philosophy to that of empirical, physical science. Indeed, while inflation predicts *many* different worlds, it is incompatible with *some* worlds – the theory is not completely unfalsifiable! One can hope to end the long-running arguments about its scientific merits by *calculating* the relevant likelihood terms.